\documentclass[manuscript]{aastex}

\slugcomment{Not to appear in Nonlearned J., 45.}

\shorttitle{Collapsed Cores in Globular Clusters}
\shortauthors{Djorgovski et al.}

\begin{document}


\title{ \bf On the dissipative non-minimal braneworld inflation}


\author{Kourosh
Nozari \altaffilmark{1}} \affil{Department of Physics,
Faculty of Basic Sciences,\\
University of Mazandaran,\\
P. O. Box 47416-95447, Babolsar, IRAN} \and
\author{M.
Shoukrani \altaffilmark{2}} \affil{Department of Physics,
Faculty of Basic Sciences,\\
University of Mazandaran,\\
P. O. Box 47416-95447, Babolsar, IRAN}


\altaffiltext{1}{knozari@umz.ac.ir}
\altaffiltext{2}{m.shoukrani@stu.umz.ac.ir}


\begin{abstract}
We study the effects of the non-minimal coupling on the dissipative
dynamics of the warm inflation in a braneworld setup, where the
inflaton field is non-minimally coupled to induced gravity on the
warped DGP brane. A warped DGP scenario is a hybrid model containing
both DGP and RSII character. We study with details the effects of
the non-minimal coupling and dissipation on the inflationary
dynamics on {\it the normal DGP branch} of this hybrid scenario in
the high-dissipation and high-energy regime. We show that
incorporation of the non-minimal coupling in this setup decreases
the number of e-folds relative to the minimal case. We also compare
our model parameters with recent observational data.\\
\end{abstract}


\keywords{Warm Inflation, Braneworld Scenario, Scalar-Tensor
Theories\\
PACS: 98.80.Cq,\, 98.80.-k,\, 04.50.-h}



\section{Introduction}

Standard big bang cosmology, with its great successes in
confrontation with observational data, has several shortcomings,
part of which can be explained naturally in inflation paradigm.
Inflation is also a successful scenario for production and evolution
of the perturbations in primary stages of the universe evolution
\citep{Lid00}. While inflationary scenario is successful in these
respects, there is a problem for realization of this scenario that
we do not know how to integrate it with ideas of particle physics
\citep{Bra05,Lid97}. From a thermodynamic viewpoint, there are two
dynamical realizations of inflation: In the standard inflation
scenario known as \emph{supercooled inflation}, radiation is
red-shifted during expansion and a vacuum dominated universe is the
result of this exponential expansion. This picture gives an
isentropic perspective of the inflation paradigm. In this picture,
the universe expands in inflation phase and its temperature decrease
rapidly. When inflation ends, a reheating period introduces
radiation into the universe. The fluctuations during this inflation
phase are zero-point ground state fluctuations and evolution of the
inflaton field is governed by the ground state evolution equation.
In this model, there are no thermal perturbations and therefore,
density perturbations are adiabatic. To have a radiation dominated
universe at the end of this inflationary phase, a reheating period
is needed to fill the universe with radiation. This model separates
expansion and reheating into two distinguished time periods.
However, energy transfer from potential energy to radiation is a
nontrivial aspect of this supercooled inflation. As a second
alternative, {\it warm inflation} proposed firstly by Berrera, is a
successful scenario to overcome this difficulty
\citep{Ber95,Ber99,Bel99a,Ber00,Bel99b, Bra03,Hal04a,Gup06,
Hal04b,Ber05a,Bas07,Ber06,Ber09,Mos08,Zha09,Rom09,del10,Mat10,
Bue11,Cai11,Bas11}. In this scenario, due to inflaton interactions
with other fields, dissipative effects arise so that radiation
production occurs concurrently with exponential expansion. Several
mechanisms for implementing such a dissipation during inflation have
been proposed \citep{Ber99,Ber01,Dym04,Ber03,Ber05b,Hal05}. For
instance, a permanent or temporary coupling of the scalar field with
other fields can lead to dissipative effects producing entropy at
different eras of inflationary stage. In warm inflation proposal,
matter and radiation energy fluctuations, which are responsible for
temperature fluctuations, are treated separately. Matter fields
interact with particles that are in a thermal bath with mean
temperature smaller than grand unified theories (GUT) critical
temperature \citep{Bel01}. Although this scenario is successful
enough, it suffers from difficulties such as unjustified initial
thermal bath which is needed to be introduced in the framework of
chaotic inflation. One important feature of the inflationary
paradigm is the fact that inflaton can interact with other fields
such as gravitational sector of the theory. This interaction is
shown by the non-minimal coupling of the inflaton field and Ricci
scalar in the spirit of the scalar-tensor theories. In fact, there
are several compelling reasons for inclusion of an explicit
non-minimal coupling of the inflaton field and gravity in the
action. For instance, non-minimal coupling arises at the quantum
level when quantum corrections to the scalar field theory are
considered. Even if for the classical, unperturbed theory this
non-minimal coupling vanishes, it is necessary for the
renormalizability of the scalar field theory in curved space. In
most theories used to describe inflationary scenarios, it turns out
that a non-vanishing value of the coupling constant cannot be
avoided. Also, in general relativity, and in all other metric
theories of gravity in which the scalar field is not part of the
gravitational sector, the coupling constant necessarily assumes the
value of \, $\frac{1}{6}$\,
\citep{Far96,Noz07a,Far00,Bou05,Noz07b,Fak90,Noz08a,Uza90,Noz09,Tsu00,
Koh05,Tsu04,Noz08b,Kof03,Lin11}. Thus, it is natural to study an
extension of the warm inflation proposal that contains explicit
non-minimal coupling of the scalar field and gravity. We call this
extension as {\it non-minimal warm inflation}. Although this issue
has been considered by some authors, but there are very limited
number of studies in this direction \citep{Bel02}. So, there is
enough motivation to study warm inflation with non-minimally coupled
inflaton field.

On the other hand, theories of extra spatial dimensions, in which
the observed universe is realized as a brane embedded in a higher
dimensional spacetime, have attracted a lot of attention in the last
few years. In this framework, ordinary matters are trapped on the
brane, but gravitation propagates through the entire spacetime
\citep{Ark98,Ant98,Ark99,Ran99a,Ran99b,Dva00}. The cosmological
evolution on the brane is given by an effective Friedmann equation
that incorporates the effects of the bulk in a non-trivial manner
\citep{Bin00,Maa10}. Other extensions of the warm inflation proposal
to braneworld scenario have been studied by some authors
\citep{Mae03,Ant07}.

With these preliminaries, the goal of the present study is to
investigate a braneworld viewpoint of warm inflation with an
explicit non-minimal coupling of the scalar field and Ricci
curvature on the brane. We study possible impact of the non-minimal
coupling and dissipation on the dynamics of these
braneworld-inspired warm inflation. Our setup is based on warped DGP
braneworld model in the platform constructed in \citep{Mae03,Ant07},
but we incorporate an explicit nonminimal coupling of the scalar
field and gravity on the brane. We study parameter space of each
model in the high dissipation limit. As we will show, this model
provides a natural exit from inflationary phase without adopting any
additional mechanism. We show also that incorporation of the
non-minimal coupling in this setup decreases the number of e-folds
relative to the minimal case. Confrontation with combined
WMAP7+BAO+$H_0$ dataset shows that this model in the high
dissipation limit mostly
gives a red and nearly scale invariant power spectrum.\\
Through this paper a dot on a quantity marks its time
differentiation, while a prime denotes differentiation with respect
to the scalar field, $\varphi$.

\section{The Setup}

We start with the following action to construct a braneworld
non-minimal inflation scenario
\begin{equation}
S=\frac{1}{2 \kappa_{5}^{2}}\int_{y\neq
0}d^5X\sqrt{-g^{(5)}}\big(R^{(5)}-2\Lambda_{5}\big)+
\int_{y=0}d^4x\sqrt{-g}\bigg[\frac{R}{2\kappa_{4}^2}-\frac{1}{2}g^{\mu\nu}
 \partial_\mu\varphi\partial_\nu\varphi-\frac{1}{2}\xi
 R\varphi^2-V(\varphi)-\lambda\bigg]
\end{equation}
In this action, which is written in the Jordan frame,  $X$ is
coordinate in the bulk, while $x$ shows induced coordinate on the
brane.\, $\kappa_5^2 $ is the $5$-dimensional gravitational
constant, $R^{(5)}$ is 5-dimensional Ricci scalar and $\xi$ is the
nonminimal coupling (NMC) of the scalar field $\varphi$ and Ricci
curvature on the brane, \,$\Lambda_{4}=\frac{1}{2}(\Lambda_{5}+
 \kappa_{4}^2 \lambda)$\, is the effective
cosmological constant on the brane,$\Lambda_{5}$ is the
5-dimensional cosmological constant in the bulk, so that
$\Lambda_{4}=0$, at least in the early universe and $\lambda$ is the
brane tension.

We have chosen a conformal coupling of the inflaton and gravity on
the brane for simplicity. In other words, non-minimal coupling of
the scalar field and gravity on the brane is
$\alpha(\varphi)=\frac{1}{2}(1-\xi\varphi^{2})$ ( with
$\kappa_{4}^{2}=8\pi G=1$). Variation of the action with respect to
$\varphi$ leads to the equation of dynamics for scalar field on the
brane
\begin{equation}
\ddot{\varphi}+3H\dot{\varphi}+\xi R\varphi+\frac{dV}{d\varphi}=0,
\end{equation}
where $R=6(\frac{\ddot{a}}{a}+\frac{\dot{a}^{2}}{a^2})$\, for
spatially flat FRW geometry on the brane and $H=\frac{\dot{a}}{a}$
is the brane Hubble parameter. If we consider phenomenologically a
dissipation coefficient $\Gamma$ that is responsible for the decay
of the scalar field into radiation during the inflationary regime,
equation of dynamics for the scalar field will be as follows
\begin{equation}
\ddot{\varphi}+3H\dot{\varphi}+\xi
R\varphi+\frac{dV}{d\varphi}=-\Gamma \dot{\varphi}
\end{equation}
$\Gamma$ can be assumed to be a constant or a function of the scalar
field $\varphi$ or the temperature $T$, or both of them. We should
emphasize that usually the dissipative coefficient is a function of
the temperature of the thermal plasma because the dissipative effect
arises from the relaxation process of the thermal plasma and
vanishes at zero-temperature. For the case where the dissipative
coefficient depends on the temperature of the thermal plasma, we
refer to
Ref.\citep{Ber95,Ber99,Bel99a,Ber00,Bel99b,Bra03,Hal04a,Gup06,
Hal04b,Ber05a,Bas07,Ber06,Ber09,Mos08,Zha09,Rom09,del10,Mat10,
Bue11,Cai11,Bas11}. In which follows for simplicity we assume that
$\Gamma$ is only a function of $\varphi$ which itself is a function
of the cosmic time. By the second Law of thermodynamics, $\Gamma$
must satisfies the condition $\Gamma=f(\varphi)>0$.

In the slow-roll approximation where $\ddot{\varphi}\ll
|3H\dot{\varphi}|$, equation of motion for the scalar field takes
the following form
\begin{equation}
\dot{\varphi}=-\frac{\xi R\varphi+V'(\varphi)}{\Gamma+3H}.
\end{equation}
The energy density and pressure of the non-minimally coupled scalar
field are given by
\begin{equation}
\rho_\varphi=\frac{1}{2}\dot{\varphi}^2+V(\varphi)+6\xi
H\varphi\dot{\varphi}+3\xi H^2\varphi^2
\end{equation}
and
\begin{equation}
P_\varphi=\frac{1}{2}\dot{\varphi}^2-V(\varphi)-2\xi
\varphi\ddot{\varphi}-2\xi \dot{\varphi}^2-4\xi
H\varphi\dot{\varphi}-\xi(2\dot{H}+3H^2)\varphi^2
\end{equation}
respectively\footnote{See the papers by Faraoni in Ref.
\citep{Far96,Noz07a,Far00,Bou05,Noz07b,Fak90,Noz08a,Uza90,Noz09,Tsu00,
Koh05,Tsu04,Noz08b,Kof03,Lin11} for discussion on the various
representations of the energy-momentum tensor of a non-minimally
coupled scalar field.}. The conservation equation for scalar field
energy density in this setup is given by
\begin{equation}
\dot{\rho}_\varphi+3H(\rho_\varphi+P_\varphi)=\dot{\varphi}V'+
\dot{\varphi}\ddot{\varphi}+3H\dot{\varphi}^2+\xi
R\varphi\dot{\varphi}
\end{equation}
where from equation (3) we find
\begin{equation}
\dot{\rho}_\varphi+3H(\rho_\varphi+P_\varphi)=-\Gamma
\dot{\varphi}^2
\end{equation}
During inflaton evolution, dissipation leads to production of
entropy. The entropy density of the radiation $S(\varphi,T)$ in this
non-minimal setup is defined by $S=-\int(\frac{dV(\varphi)}{T})$.
Now the total energy density of the system is defined by
\begin{equation}
\rho=\frac{1}{2}\dot{\varphi}^2+V(\varphi)+6\xi
H\varphi\dot{\varphi}+3\xi H^2\varphi^2+TS
\end{equation}
and using relation (8) the rate of entropy production can be
deduced. In this non-minimal setup, entropy production is affected
by the non-minimal coupling of gravity and inflaton field. This can
be seen more explicitly in the following relation
\begin{equation}
T(\dot{S}+3HS)=\Gamma\dot{\varphi}^{2},
\end{equation}
where by equation (4),\, $\dot{\varphi}$ is directly related to the
non-minimal coupling. The basic idea of warm inflation is that
radiation production is occurring concurrently with inflationary
expansion due to dissipation from the inflaton field system. The
equation of state for radiation is given by
$P_\gamma=\frac{\rho_{\gamma }}{3}$. Therefore, the conservation
equation of $\rho_\gamma$ yields the following result
\begin{equation}
\dot{\rho}_\gamma+4H\rho_\gamma=\Gamma\dot{\varphi}^2.
\end{equation}
It is interesting to note that time evolution of the radiation field
is related to the non-minimal coupling via $\dot{\varphi}$.
Following \citep{Mae03,Ant07}, we assume a quasi-stable radiation
production during the warm inflation phase, namely we assume that
$\dot{\rho}_\gamma\ll 4H\rho_{\gamma}$ and
$\dot{\rho}_\gamma\ll\Gamma\dot{\varphi}^2$, so we can write
\begin{equation}
\rho_\gamma\approx\frac{\Gamma}{4H}\dot{\varphi}^2.
\end{equation}
In the slow-roll approximation, $\rho_\varphi$ and $\rho_\gamma$
attain the following approximate forms respectively
\begin{equation}
\rho_\varphi\approx V(\varphi)+6\xi H\varphi\dot{\varphi}+3\xi
H^2\varphi^2,
\end{equation}
\begin{equation}
\rho_\gamma\approx 0.
\end{equation}
we use these approximate forms in equation (17)to find cosmological
dynamics of the model . We define the dissipation factor as follows
\begin{equation}
Q\equiv\frac{\Gamma}{3H},
\end{equation}
which is a dimensionless parameter. With this definition, we can
write
\begin{equation}
\dot{\varphi}=-\frac{\xi R\varphi+V'(\varphi)}{3H(1+Q)}
\end{equation}
During the slow-roll stage, the scalar field evolution is damped.
For the high (or weak) dissipation regime we have $Q\gg1$(or
$Q\ll1$) respectively. Now the Friedmann equation of the model takes
the following form ( for details of calculations see the paper by
Nozari and Fazlpour in
\citep{Far96,Noz07a,Far00,Bou05,Noz07b,Fak90,Noz08a,Uza90,Noz09,Tsu00,
Koh05,Tsu04,Noz08b,Kof03,Lin11} which considers a general
non-minimal Gauss-Bonnet scenario with induced gravity. See also the
paper by Kofinas {\it et al.} in
\citep{Far96,Noz07a,Far00,Bou05,Noz07b,Fak90,Noz08a,Uza90,Noz09,Tsu00,
Koh05,Tsu04,Noz08b,Kof03,Lin11})
\begin{equation}
H^2+\frac{k}{a^{2}}-\frac{{\cal{E}}_{0}}{a^{4}}=\Bigg[r_{c}\alpha(\varphi)\big(H^{2}+\frac{k}{a^{2}}\big)-
\frac{\kappa_{5}^2}{6}\big(\rho+\lambda \big)\Bigg]^{2}
\end{equation}
where we assume that the metric on the brane is spatially flat FRW
type $(k=0)$. \,$r_{c}=\frac{\kappa_{5}^{2}}{2\kappa_{4}^{2}}$\,, is
the induced-gravity crossover length scale, $\rho$ is the total
energy density on the brane, namely $\rho=\rho_\varphi+\rho_\gamma$.
Also ${\cal{E}}_{0}$ is a constant related to the 5-dimensional Weyl
tensor describing the effect of the bulk graviton degrees of freedom
on the brane dynamics and corresponding term is called dark
radiation term. Note that the bulk cosmological constant appears
implicitly via $\lambda$ that
\,$\Lambda_{4}=\frac{1}{2}(\Lambda_{5}+
 \kappa_{4}^2 \lambda)$\,.\\
Note that in (17) if we consider a flat brane $(k=0)$ with
$\lambda=0$ and also a Minkowski bulk
$(\Lambda_{5}=0,{\cal{E}}_{0}=0)$, then we recover the pure DGP
solution as follows
\begin{equation}
 H^{2}=\pm \frac{H}{r_{c}\alpha(\varphi)}+\frac{{\kappa_{4}^{2}}}{{3\alpha(\varphi)}}\rho.
\end{equation}
In this work we consider the case with  $\Lambda_{4}$.
\begin{equation}
H^2-\frac{{\cal{E}}_{0}}{a^{4}}=\Bigg[r_{c}\alpha(\varphi)H^{2}-\frac{\kappa_{5}^2}{6}\bigg(V(\varphi)-\frac{2\xi^{2}R\varphi^{2}+2\xi\varphi
 V'}{1+Q}+3\xi H^{2}\varphi^{2}+\lambda \bigg)\Bigg]^{2}
\end{equation}
Using the definition of $\dot{\varphi}$ as given by equation (16),
this leads to a forth order equation for Hubble parameter of the
model$$H^{4}\Big(r_{c}\alpha(\varphi)-\frac{\xi
\varphi^{2}\kappa_{5}^{2}}{2}\Big)^{2}-H^{2}\Bigg[1+2\Big(r_{c}\alpha(\varphi)-\frac{\xi
\varphi^{2}\kappa_{5}^{2}}{2}\Big)\Bigg(\frac{\kappa_{5}^2}{6}\Big(V(\varphi)-
\frac{2\xi^{2}R\varphi^{2}+2\xi\varphi
 V'}{1+Q}+\lambda \Big)\Bigg)\Bigg]
$$
\begin{equation}
+\Bigg[\frac{\kappa_{5}^2}{6}\Big(V(\varphi)-\frac{2\xi^{2}R\varphi^{2}+2\xi\varphi
 V'}{1+Q}+\lambda
 \Big)\Bigg]^{2}+\frac{{\cal{E}}_{0}}{a^{4}}=0.
\end{equation}By assuming a non-vanishing non-minimal coupling, that is if
$\xi\neq 0$, we can solve equation (20) for $H^2$ to find the
following solutions
\begin{equation}
H^2=\frac{\lambda}{b^2}\,\Big[\,\frac{8\pi}{m_{p}^2}+\frac{8\pi
r_c}{3 m_{p}^2}
  \,b\,(1+\frac{d}{\lambda})\,\Big]\pm\frac{2}{b}\,\sqrt{\frac{1}{b^2}+\frac{r_c}{b}(1+\frac{d}{\lambda})-\frac{{4\cal{E}}_{0}}{a^{4}}}
\end{equation}
where $d$ and $b$ are defined as follows
\begin{equation}
d\equiv V-\frac{2\xi\varphi}{1+Q}(\xi R\varphi+V'),\quad\quad
b\equiv r_c\alpha(\varphi) - \frac{3\xi\varphi^2 \kappa_{5}^{2}}{2}.
\end{equation}
Since we are interested in the inflationary dynamics of the model,
we will neglect the dark radiation term in which follows.In this
case, generalized Friedmann equation (21) attains the following form
\begin{equation}
H^2=\frac{\lambda}{b^2}\,\Big[\,\frac{8\pi}{m_{p}^2}+\frac{8\pi
r_c}{3 m_{p}^2}
  \,b\,(1+\frac{d}{\lambda})\,\Big]\pm\frac{2}{b}\,\sqrt{\frac{1}{b^2}+\frac{r_c}{b}(1+\frac{d}{\lambda})}
\end{equation}
Note that the effective cosmological constant on the brane is
determined by $\Lambda_{4}= \frac{4\pi}{m_{5}^3}
\big(\Lambda_{5}+\frac{4\pi}{3m_{5}^3}\lambda^{2}\big)$ and the
four-dimensional Plank scale is given by
$m_{p}=\sqrt{\frac{3}{4\pi}}(\frac{m_{5}^3}{\sqrt{\lambda}})m_{5}$ .
To be more specific, in which follows we consider just the minus
sign in equation (23). Now, we define the slow-roll parameters as
follows
$$\varepsilon\equiv-\frac{\dot{H}}{H^2},$$
$$\eta\equiv-\frac{\ddot{H}}{H\dot{H}}$$
and
$$\gamma^{2}\equiv2\varepsilon\eta-\frac{d\eta}{dt}$$
We note that in the analysis of the warm inflation, the slow-roll
conditions for thermal plasma should be considered too. In this
respect we require that in equation (12) one should impose a
condition for suppression of $\ddot{\rho_{\gamma}}$ (see for
instance \citep{Ber95,Ber99,Bel99a,Ber00,Bel99b,Bra03,Hal04a,Gup06,
Hal04b,Ber05a,Bas07,Ber06,Ber09,Mos08,Zha09,Rom09,del10,Mat10,
Bue11,Cai11,Bas11} for details). Since in our analysis we have not
included the temperature dependence of the scalar field potential,
the forth slow-roll parameter defined for thermal plasma is absent.
We emphasize that the slow-roll approximation is valid when all of
the slow-roll parameters are smaller than $1 + Q$.

We need to calculate these parameters in our non-minimal dissipative
braneworld model. For $\varepsilon$ we find
\begin{equation}
\varepsilon\approx\frac{2\dot{A}B -2\dot{B}A -\dot{C} C^{-1}
A}{4H^{3} A},
\end{equation}
where
$$ A\equiv\frac{\Big(r_c\alpha(\varphi) - \frac{3\xi\varphi^2 \kappa_{5}^{2}}{2}\Big)^{2}}{\lambda}$$
$$B\equiv\frac{8\pi}{m_{p}^2}+\frac{8\pi r_c}{3 m_{p}^2}\, b\,(1+\frac{d}{\lambda}),$$
$$C\equiv\frac{4}{b^4}+\frac{4 r_c}{b^3}\,(1+\frac{d}{\lambda})$$
and  $d$ and $b$ are defined in equation (22). The second slow-roll
parameter, $\eta$, is calculated as follows
$$\eta \approx\bigg[4A^{2}H^{3}(2\dot{B}A
-2\dot{B}\dot{A}+\dot{C}C^{-1}A)\bigg]^{-1}\bigg[(8 \dot{A} A H^{2}
+2\dot{B} A-2\dot{A} B+\dot{C}C^{-1}A)(2\dot{B}A -2\dot{A}B +\dot{C}
C^{-1}A)
 $$
\begin{equation}
-(2\ddot{B} A-2\ddot{A}B + \ddot{C} C^{-1} A-\dot{C}^{2}C^{-2}A
+\dot{C} C^{-1} \dot{A}) 4A^{2}H^{2}\bigg].
\end{equation}
And finally, the third slow-roll parameter, $\gamma^2$, in our model
takes the following form
$$\gamma^2\approx-\frac{\dot{\varphi}}{(1+Q)H}\bigg(\frac{V'''}{3H^2}-\frac{Q'(V''+\xi
    R)}{3H^2(1+Q)}+Q '\varepsilon\bigg)$$
\begin{equation}
-\frac{\dot{\varphi}^2}{(1+Q)H^2}\Big(Q''-\frac{Q'^2}{1+Q}\Big)-\frac{Q'(\xi
R\varphi+V')}{3(1+Q)^2H^2}\Big(\eta+\varepsilon-\frac{2Q'\dot{\varphi}}{(1+Q)H}\Big).
\end{equation}
After constructing the basic formalism, we study inflationary
dynamics of this non-minimal dissipative model. For this goal, in
which follows and in all of our numerical calculations, we consider
the well-known large-field inflationary potential $V(\varphi)=V_0
\varphi^n$ to investigate outcomes of our model. The inflationary
phase terminates if the condition $\varepsilon=1+Q$ is satisfied and
this happens when radiation and potential energies satisfy an
especial condition will be derived later on ( see equation (31)).
Figure $1$ shows the variation of $\varepsilon$ versus the inflaton
field in large field limit with $V(\phi)=V_{0}\varphi^{2}$. In this
figure we have assumed high-dissipation and high energy limit with
$Q\gg 1$ and $V\gg \lambda$ respectively. As this figure shows, this
model has the capacity to accounts for natural exit from
inflationary phase without any additional mechanism. In fact,
non-minimal coupling of the inflaton field and gravity on the brane
provides a suitable parameter space for natural exit of the
inflationary phase. We stress that in this figure the minimal case
with $\xi=0$ accounts for natural exit without additional mechanism
too. This is possible in our setup because of the braneworld effect.
We note that most inflationary models exit inflation naturally, e.g.
chaotic, natural new inflation etc. But there are scenarios such as
hybrid inflation that introduce a phase transition to terminate
inflation. Figure $2$\, shows the possibility of the natural exit
from inflationary phase in high-dissipation ($Q\gg1$) and
high-energy limit ($V\gg \lambda$) for
$V(\phi)=V_{0}\exp{\big(-\sqrt{\frac{2}{p}}\frac{\varphi}{m_{p}}}\big)$.
We see that in this case incorporation of the non-minimal coupling
and braneworld effects are sufficient for natural exit from
inflationary phase too and there is no need to introduce additional
mechanism to achieve this goal.
 Now the energy density of the radiation field as a function of the scalar field potential can be
written as follows
\begin{equation}
\rho_\gamma=\frac{3Q}{4}\bigg(\frac{\xi
R\varphi+V'}{3(1+Q)}\bigg)^2\frac{\varepsilon}{\zeta},
\end{equation}
where $\zeta\equiv\varepsilon H^{2}$. Since $\rho_{\gamma}=\sigma
{T_{\gamma}}^{4}$, where $\sigma$ is the Stefan-Boltzmann constant
and $T_{\gamma}$ is the temperature of the thermal bath, we find
\begin{equation}
T_{\gamma}=\Bigg[\frac{3Q}{4\sigma}\bigg(\frac{\xi
R\varphi+V'}{3(1+Q)}\bigg)^2\frac{\varepsilon}{\zeta}\Bigg]^{1/4}.
\end{equation}
On the other hand, the relation between $\rho_{\gamma}$ and
$\rho_{\varphi}$ is given by
\begin{equation}
\rho_\gamma=\bigg[\frac{3Q}{4}\Big(\frac{\xi
R\varphi+V'}{3(1+Q)}\Big)\frac{\varepsilon}{\zeta}\bigg]\bigg
(\frac{V-\rho_\varphi+3\xi H^2\varphi^2}{2\xi\varphi}\bigg).
\end{equation}
Inflation occurs when the condition $\varepsilon<1+Q$ (or
$\ddot{a}>0)$ is fulfilled. Therefore, using (28) we find
\begin{equation}
\rho_\varphi<-\bigg(\frac{2\xi\varphi}{\frac{Q}{4\zeta}(\frac{\xi
 R\varphi+V'}{3(1+Q)})}\bigg)\rho_\gamma+V+3\xi H^{2}\varphi^2,
\end{equation}
which is the condition for realization of the non-minimal warm
inflation on the brane. Note that $\zeta$ is a function of the
3-brane tension $\lambda$, and inflaton potential, $V$. Therefore,
to have a warm inflationary period on the brane, the condition (30)
should be fulfilled. As we have pointed out previously, inflationary
phase will terminates when the universe heats up so that the
condition $\varepsilon= 1+Q $ is satisfied. In our non-minimal case
this implies that
\begin{equation}
 \rho_\varphi\simeq-\bigg(\frac{2\xi\varphi}{\frac{Q}{4\zeta}(\frac{\xi
   R\varphi+V'}{3(1+Q)})}\bigg)\rho_\gamma+V+3\xi\zeta\varphi^2.
\end{equation}
By comparison with the minimal case as has been studied in Ref.
\citep{Mae03,Ant07}, we see that non-minimal coupling of the scalar
field and gravity on the brane has important role on the natural
exit of the inflationary phase. This role has been highlighted in
figures $1$ and $2$.

Now we focus on the number of e-foldings as another important
quantity in a typical inflation scenrio. The number of e-foldings is
defined as
$$N\equiv\int_{t_{i}}^{t_{end}}H
dt=\int_{\varphi_{i}}^{\varphi_{end}}\frac{H}{\dot{\varphi}}d\varphi$$
In our setup, the number of e-foldings at the end of the inflation
phase is given by
\begin{equation}
N=-\int_{\varphi_i}^{\varphi_{e}}\bigg(\frac{3H^2(1+Q)}{\xi
R\varphi+V'}\bigg)d\varphi.
\end{equation}
Using the explicit form of $H^{2}$ with negative sign as given by
equation (23), the number of e-foldings in our non-minimal setup
takes the following form
\begin{equation}
N=-\int_{\varphi_{i}}^{\varphi_{e}}\bigg(\frac{3(1+Q)}{\xi
R\varphi+V'}\bigg)\Bigg\{\frac{\lambda}{b^2}\Big[\frac{8\pi}{m_{p}^2}+\frac{8\pi
r_c}{3 m_{p}^2}
  \,b\,(1+\frac{d}{\lambda})\Big]-\frac{2}{b}\,\sqrt{\frac{1}{b^2}+\frac{r_c}{b}(1+\frac{d}{\lambda})}\Bigg\}d\varphi
\end{equation}
with $d$ and $b$ as defined in (22).\,  $\varphi_{i}$  denotes the
value of the scalar field $\varphi$ when universe scale observed
today crosses the Hubble horizon during inflation, while
$\varphi_{e}$ is the value of the scalar field when the universe
exists the inflationary phase. The role played by the dissipation
factor is the same as the minimal case studied in
\citep{Mae03,Ant07}: in the high-energy and high-dissipation limit
where $V\gg\lambda$ and $Q\gg1$ respectively, the rate of expansion
is increased in comparison with the case in standard inflationary
model, while in the low energy limit and weak dissipation where
$V\ll\lambda$ and $Q\ll 1$ respectively, the rate of expansion is
nearly the same as standard inflationary model. In our non-minimal
case, the role played by the non-minimal coupling is not so trivial
due to complicated form of the equation (33). To treat this problem,
we define
$$
g\equiv-\bigg(\frac{3(1+Q)}{\xi
R\varphi+V'}\bigg)\Bigg\{\frac{\lambda}{b^2}\Big[\frac{8\pi}{m_{p}^2}+\frac{8\pi
r_c}{3 m_{p}^2}
  \,b\,(1+\frac{d}{\lambda})\Big]-\frac{2}{b}\,\sqrt{\frac{1}{b^2}+\frac{r_c}{b}(1+\frac{d}{\lambda})}\Bigg\}
$$
The number of e-foldings is the area enclosed between $g$-curve and
the horizontal axis from $\varphi_{i}$ to $\varphi_{e}$ in the
$g-\varphi$ plane. Figure $3$ shows the variation of $g$ versus
$\varphi$ for different values of the non-minimal coupling for
$\Gamma=\Gamma_{0}\varphi^{2}$. In the high and low energy limits,
the number of e-folds decreases by increasing the values of $\xi$ (
in the case that all other quantities are fixed). Therefore,
incorporation of the non-minimal coupling in this setup essentially
decreases the number of e-folds. In an approximate manner, if we
assume that $\xi$ is a small parameter ( see for instance
\citep{Noz08c,Noz10} to see viability of this assumption), in the
low-energy ($V\ll\lambda$) and low-dissipation limit ($Q\ll 1$) we
can ignore terms containing $\frac{d}{\lambda}$ in the equation
defining the number of e-folds and the remaining expression leads to
the result that number of e-foldings decreases by inclusion of the
non-minimal coupling. We stress that this decreasing depends on the
values of the non-minimal coupling, inflation potential and also the
detailed form of the dissipative coefficient. In the high energy and
high dissipation limit with $V\gg\lambda$ and $Q\gg1$ respectively,
the number of e-folds decreases by inclusion of the non-minimal
coupling too.
 Now,we study scalar and tensor perturbations in this non-minimal
dissipative model. In the case of scalar perturbations, the scalar
and the radiation fields are interacting. These perturbations are
supposed to be adiabatic on the brane. Dissipative effects can
produce a variety of spectral index, ranging between red and blue,
and thus producing the running blue to red spectral index suggested
by WMAP data \citep{Kom08,Spe07,Jar10}.We note that in the
background spacetime $\frac{{\cal{E}}_{0}}{a^{4}}=0$, and we can use
equation (23) as Friedmann equation in this setup. But the perturbed
FRW brane has a nonzero $\frac{{\cal{E}}_{0}}{a^{4}}$, which encodes
the effects of the bulk gravitational field on the brane
\citep{Koy06} and we have use the Friedmann equation (21). We define
the scalar curvature perturbation amplitude of a given mode when
re-enters the Hubble radius as follows
\begin{equation}
\delta_H=\frac{2}{5} \frac{H}{\dot{\varphi}}\delta\varphi.
\end{equation}
We note that this expression is actually for 4-dimensional case; but
as Maartens {\it et al.} have shown in Ref. \citep{Mar00}, it can be
used in this braneworld setup too. Note that although we have used
the usual spectrum for the field perturbations in this warm
inflation scenario, but the presence of the non-minimal coupling is
implicit in the form of Hubble parameter, $H$, in this setup. Given
that this spectrum have been derived in a complete different set-up,
to begin with without coupling of the scalar field to the Ricci
scalar, it may hold in the scenario studied in this work if we use
the form of $H$ dependent on the non-minimal coupling.

The scalar spectrum index is defined as follows
\begin{equation}
 n_s=1+\frac{d\ ln \delta_H^2}{d\ ln k}
\end{equation}
The interval in wave number is related to the number of e-foldings
by the relation $d \ln k(\varphi)= -d N(\varphi)$. Substituting (16)
into the relation (34) for $\delta_H$, we find ( we refer to the
Ref. \citep{Noz11} for detailed form of fluctuations in the warm
inflation scenarios)
\begin{equation}
\delta_H^2=\frac{36}{25}\frac{H^4(1+Q)^2}{(\xi
 R\varphi+V')^2} \delta\varphi^2
\end{equation}
Using the slow-roll parameters, we have
\begin{equation}
n_s=1-\frac{7}{2}\varepsilon+
   \frac{3}{2}\eta-\frac{9Q'\dot{\varphi}^2}{4(\xi R\varphi+V')}
\end{equation}
The running of the spectral index which is defined as
\begin{equation}
\alpha_{s}=\frac{dn_{s}}{d\ln k}
\end{equation}
in our model takes the following form
$$\alpha_s=\frac{9}{2}(2\varepsilon\eta-4\varepsilon^2)+\frac{3}{2}\Big(2\varepsilon\eta-\gamma^2\Big)
-\frac{(\xi R\varphi+V')}{12(1+Q)^3H^4}\bigg[Q''(\xi
R\varphi+V')+Q'(\xi R+V'')\bigg]$$
\begin{equation}
 +\frac{Q'(\xi R\varphi+V')}{2(1+Q)^3H^2}\bigg(\frac{Q'(\xi
 R\varphi+V')}{3H^2(1+Q)}-(1+Q)\varepsilon\bigg).
\end{equation}
Now we study these quantities numerically to investigate
implications of this non-minimal dissipative setup. Figure $4$ shows
the variation of $n_{s}$ versus the inflaton field for different
values of the non-minimal coupling in high-dissipation limit for
$V=V_0\varphi^2$. Since $n_s\leq 1$, our non-minimal dissipative
model favors a red power spectrum. Therefore, our model for this
type of potential gives also a nearly scale invariant spectrum. The
result of WMAP7 for $\Lambda$CDM gives
$n_{s}=0.963^{+0.014}_{-0.014}$\, for index of the power spectrum
\citep{Kom08,Spe07,Jar10}. Combining WMAP7 with other data sources (
 the Baryon Acoustic Oscillations (BAO) and$H_{0}$), gives
$n_{s}=0.963^{+0.012}_{-0.012}$ \citep{Kom08,Spe07,Jar10}. These
results show that a red power spectrum is favored and $ n_{s} > 1$
is disfavored even when gravitational waves are included, which
constrains the models of inflation that can produce significant
gravitational waves, such as chaotic or power-law inflation models,
or a blue spectrum, such as hybrid inflation models. So, our model
at least qualitatively is in agreement with recent observational
data. If there is running of the spectral index, the constraints on
the spectral index and its running are given by
$n_{s}=1.017^{+0.042}_{-0.043}$ \, and $\frac{dn_{s}}{d\ln
k}=-0.022^{+0.020}_{-0.020}$ respectively \citep{Kom08,Spe07,Jar10}.
The special case with $n_{s} = 1$ and $\frac{dn_{s}}{d l\ln k} = 0$
results in the scale invariant spectrum. The significance of $n_{s}$
and $\frac{dn_{s}}{d l\ln k}$ is that different inflation models
motivated by different physics make specific, testable predictions
for the values of these quantities. In summary, we see that our
dissipative non-minimal model mostly excludes blue spectrum in
agreement with recent observational data.
 Figure $5$ shows the running of the spectral index versus the inflaton field in
high-dissipation limit for $V=V_0\varphi^2$. As we see, there is no
signature of a positive running of the spectral index in this setup
with positive values of the non-minimal coupling. Note that a
negative $\xi$ is theoretically interesting but it requires
anti-gravitation and therefore has been excluded in our analysis.
Now we pay attention to the tensorial perturbations. As it has been
mentioned in Ref.\citep{Bha06}, the generation of the tensor
perturbations during inflation produces stimulated emission in the
thermal background of gravitational waves. This process changes the
tensor modes by an extra temperature dependent factor given by
$\coth(\frac{k}{2T})$\, where $T$ is the temperature of the thermal
bath. The corresponding spectrum becomes
\begin{equation}
A_g^2=\frac{16\pi}{\mu^2}(\frac{H}{2\pi})^2\coth[\frac{k}{2T}],
\end{equation}
where $H$ is given by equation (21) with negative sign. Note that we
incorporated the bulk effects in this relation via the Hubble
parameter of model which contains especially the effect of bulk
radiation term. Using equation (36) and (40), in the limit of
$Q\gg1$ the tensor to scalar ratio is given by
\begin{equation}
r=\bigg(\frac{A_g^2}{P_{R}}\bigg)_{k=k_{i}}\simeq\frac{8\varphi}{9\sqrt{3}\mu^2}\Bigg[\frac{(\xi
R\varphi+V'(\varphi))}{TH^3Q^{5/2}}\coth(\frac{k}{2T})\Bigg]_{k=k_{i}}.
\end{equation}
Here, $P_{R}=\frac{25}{4} \delta_H^{2}$ and $k_i$ is referred to
$k=Ha$, the value of $k$ when the universe scale crosses the Hubble
horizon during inflation. Our model with $V_{0}=1$,\, $
m_{p}=10^{19}$ GeV,\, $Q=10^{4}$ and $T=2.4 \times 10^{17}$ GeV
gives $r< 0.24$ which is comparable with the observationally
supported value of $r< 0.22$ from WMAP7+BAO+$H_{0}$ combined
dataset. We note that the above choices of parameters to calculate
scalar-to-tensor ratio, generate suitable amplitude of scalar
perturbation well in the range of observationally supported value of
$P_{R} \sim 5\times 10^{-5}$.

Finally, a point should be stressed about the issue of frames in our
setup. There is a conformal transformation which transforms the
action (1) ( which is written in the Jordan frame) to corresponding
action in the Einstein frame. In the Einstein frame, the
gravitational sector is expressed in terms of a re-scaled scalar
field which is minimally coupled and evolves in a re-scaled
potential, thereby simplifying the formalism. However, one has to
keep in mind that the matter sector is strongly affected by such a
conformal transformation since all of the matter fields are now
non-minimally coupled to the re-scaled metric: in particular, stress
tensor conservation in the matter sector is no longer ensured
\citep{Sch05}. On the other hand, as Makino and Sasaki \citep{Mak91}
and Fakir {\it et al} \citep{Fak92} have shown, the amplitude of
scalar perturbation in the Jordan frame exactly coincides with that
in the Einstein frame. This proof (for details see \citep{Kom99})
allows us to calculate the scalar power spectrum in the Jordan and
Einstein frame. As a result, the scalar power spectrum has no
dependence on the choice of frames, i.e., it is conformally
invariant. So, our results can be compared to observations directly
without any ambiguities \citep{Kom99}. This is an important point
since one has to check validity of non-gravity experiments in
Einstein frame. For instance, validity of electro-magnetic related
experiments such as CMB experiment should be checked in Einstein
frame. As Komatsu and Futamase have shown, the scalar power spectrum
is independent on the choice of frames \citep{Kom99}. We terminate
this section by pointing that from a statistical mechanics
viewpoint, the interaction between quantum fields and a thermal bath
in this warm inflation scenario could be illustrated by a general
fluctuation-dissipation relation ( see for instance \citep{Wei93}).
Warm inflation was criticized on the basis that the inflaton cannot
decay during the slow-roll phase \citep{Yok99}. However, in recent
years, it has been shown that the inflaton can indeed decay during
the slow-roll phase and now this scenario rests on solid theoretical
grounds ( see for instance
\citep{Ber95,Ber99,Bel99a,Ber00,Bel99b,Bra03,Hal04a,Gup06,
Hal04b,Ber05a,Bas07,Ber06,Ber09,Mos08,Zha09,Rom09,del10,Mat10,
Bue11,Cai11,Bas11} and \citep{Mae03,Ant07}).

\section{Summary}
 It is well-known that early
universe was filled by radiation energy $(\rho_{\gamma})$ and vacuum
energy $(\rho_{v})$ (with equation of state as $\rho_{v}=-p_{v}$).
In the inflationary regime, with positively accelerated expansion,
$\ddot{a}>0$, one has $\rho_{v}>\rho_{\gamma}$. At the end of the
inflationary phase however, universe is radiation dominated and
therefore inflation sustains vacuum energy density along expansion
of scale factor and at the time of the end of inflation universe
enters to the radiation dominated era. This mean that after the end
of inflation the universe is radiation dominated but the inflaton
field retains some of its potential energy, because the inflaton
condensate does not need to decay to reheat inflation. In
inflationary paradigm, inflaton can interact with other fields such
as Ricci curvature. This interaction is shown by the non-minimal
coupling (NMC) of the inflaton and gravity in the action of the
scenario. There are many compelling reasons to include an explicit
non-minimal coupling in the action. So, naturally we should include
non-minimal coupling of gravity and inflaton in the action of the
inflation scenario. Usually, by incorporation of the non-minimal
coupling it is harder to realize inflation even with potential that
are known to be inflationary in the minimal case. However, inclusion
of the non-minimal coupling is inevitable from field theoretical
viewpoint especially their renormalizability. From a thermodynamical
viewpoint to the inflationary expansion, there are two dynamical
realizations of inflation. Standard picture is isentropic inflation
which is referred to as supercooled inflation. In this picture,
universe expands in inflation phase and its temperature decrease
rapidly. When inflation ends, a reheating period introduces
radiation into the universe. The fluctuations in this type of
inflation model are zero-point ground state fluctuations and
evolution of the inflaton field is governed by a ground state
evolution equations. In this type of models we have not any thermal
perturbations and therefore, density perturbations are adiabatic.
The other picture is a non-isentropic inflation or warm inflation.
Warm inflation is a model of inflationary dynamics which by adopting
it, we can release from reheating period. In warm inflation
scenario, interaction between the inflaton and other fields causes
the radiation energy density. In this picture, inflation terminates
smoothly and radiation regime is dominated, without a reheating
period. The fluctuations during inflation emerge from some excited
states and the evolution of the inflaton has dissipative terms
arising from the interaction of the inflaton with other fields. In
this picture, dissipative effects are important during the inflation
period. In warm inflation, density fluctuations arise from thermal
fluctuations and this fluctuations in the radiation cause the
entropy perturbations. This entropy fluctuations disappear before
inflation ends. For this reason, solution of the warm inflation
violates the adiabatic condition. In this model, inflaton interacts
with a thermal bath in warm inflation period by a friction term and
this causes decay of inflaton. The friction term itself enters into
the dynamics of the scalar field equation as an ad hoc term. In
fact, warm inflation predicts that inflaton interactions with
surrounding fields during the inflationary period will result in a
friction term in the equations of motion. In this scenario, the
anisotropies seen in the CMB are produced by thermal fluctuations.
As a promising feature, it may be possible to distinguish between
warm inflation and standard supercooled inflation models using
forthcoming satellite data. In this paper we have studied warm
inflation by incorporating an explicit non-minimal coupling of the
inflaton field and gravity on the normal branch of a warped DGP
brane.We have studied also parameter space of the model in high
dissipation and high energy limits. As we have shown, this model
provides natural exit from the inflationary phase without adopting
any additional mechanism(s) for appropriate values of the
non-minimal coupling. By a numerical analysis in this case, we have
shown that incorporation of the non-minimal coupling in this setup
decreases the number of e-folds relative to the minimal case. We
emphasize however that this result is sensitive to the details of
the model parameters such the dissipation function, scalar field
potential and the non-minimal coupling. A qualitative confrontation
with WMAP7+BAO+$H_{0}$ combined data shows that this model in high
dissipation limit gives a red and nearly scale invariant power
spectrum for positive values of the non-minimal coupling. From
another viewpoint, comparison of this non-minimal warm inflation
model with observational data could provide more accurate
constraints on the values of the non-minimal coupling. Since we have
calculated the first order contributions in slow-roll parameters,
our results are valid in both Jordan and Einstein frame. However,
higher order corrections could lead to different results in these
two frames.

\newpage
\begin{figure}[htp]
\begin{center}
\includegraphics{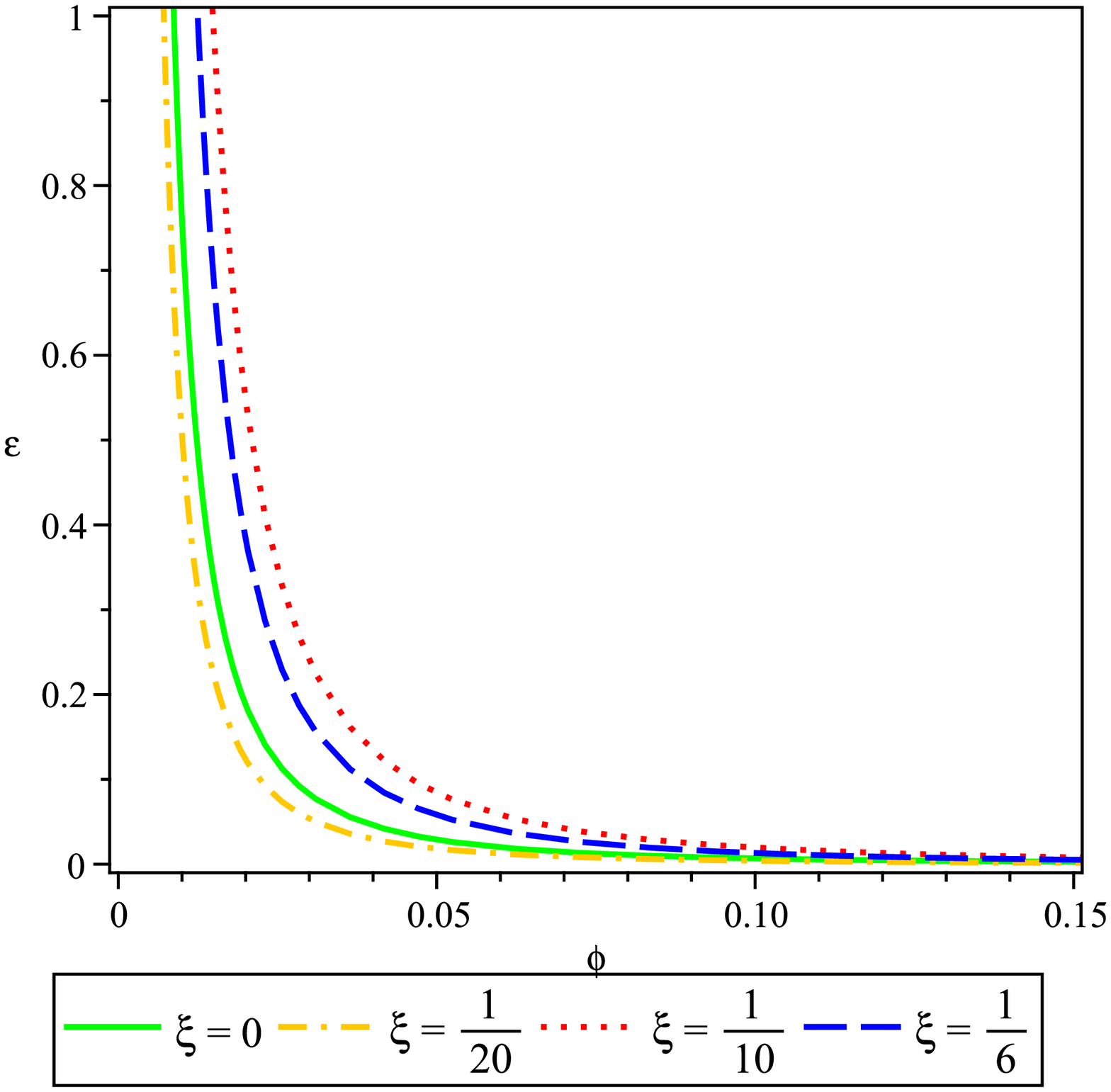}
\end{center}
\vspace{8 cm}
 \caption{\small { Variation of the $\varepsilon$ versus scalar field with
$V(\phi)=V_{0}\varphi^2$ for different values of the non-minimal
coupling in high-dissipation ($Q\gg 1$) and high-energy ($V\gg
\lambda$) limit. There is the possibility to account for the
condition $\varepsilon=1+Q$ which shows the end of the inflationary
phase. This can be achieved naturally without any additional
mechanism.}}
\end{figure}

\begin{figure}[htp]
\begin{center}
\includegraphics{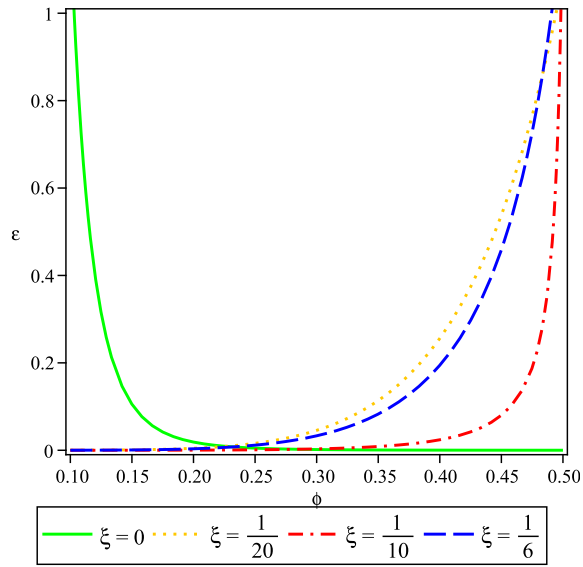}
\end{center}
\vspace{6.5 cm} \caption{\small { Variation of $\varepsilon$ versus
$\varphi$ for different values of the non-minimal coupling in
high-dissipation ($Q\gg1$) and high-energy ($V\gg \lambda$) limit
for
$V(\phi)=V_{0}\exp{\big(-\sqrt{\frac{2}{p}}\frac{\varphi}{m_{p}}}\big)$
with $p=50$ and $m_{p}=1$. The condition $\varepsilon=1+Q$ which
shows the end of the inflationary phase is achievable naturally in
this setup without any additional mechanism.}}
\end{figure}

\begin{figure}
\begin{center}\includegraphics{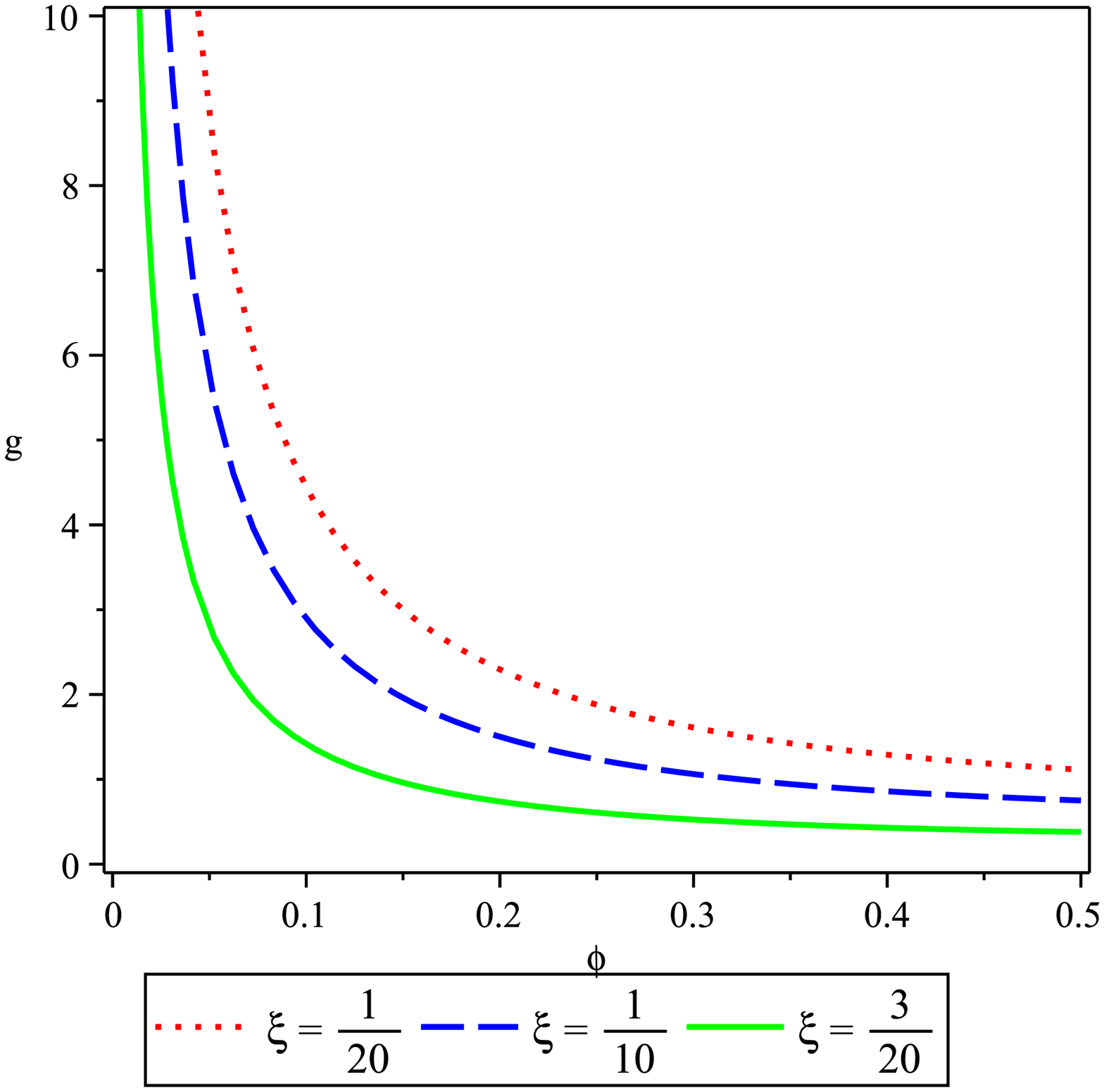} \vspace{5.5cm}\includegraphics{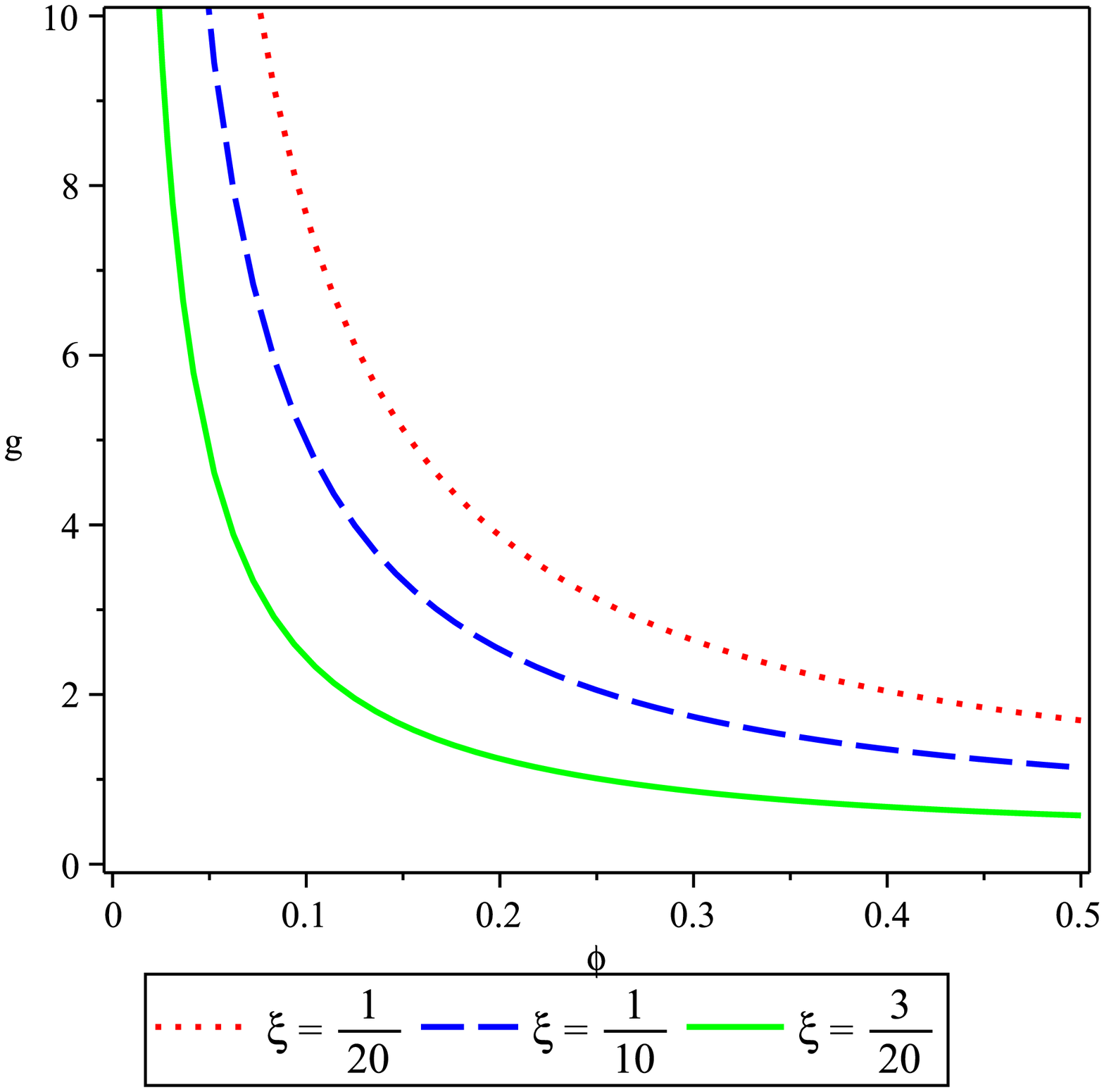}
\end{center}
\caption{\small {$g$ versus $\varphi$ in the limit of high ( left
hand side) and low energy (right hand side) limits of the scenario
with various values of the non-minimal coupling. Number of e-folds
is given by the area bounded between $g$-curve and the horizontal
axis in this $g - \varphi$ plot. In the high energy and low energy
limits, the number of e-folds decreases by increasing the values of
$\xi$. Therefore, incorporation of the non-minimal coupling in this
setup essentially decreases the number of e-folds. Nevertheless, we
stress that this decreasing depends on the values of the non-minimal
coupling, inflation potential and also the detailed form of the
dissipative coefficient. }}
\end{figure}

\begin{figure}[htp]
\begin{center}
\includegraphics{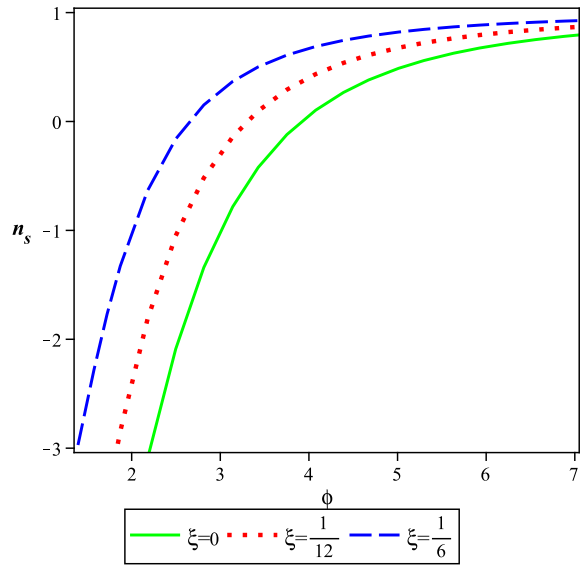}
\end{center}
\vspace{8 cm}
 \caption{\small {Variation of the spectral index $n_{s}$ versus inflaton field for different values
of the non-minimal coupling in high-dissipation limit for
$V=V_0\varphi^2$. Evidently, $n_s\leq 1$ and therefore our
non-minimal model favors a red power spectrum and this is confirmed
by recent observational data sets. }}
\end{figure}

\begin{figure}[htp]
\begin{center}
\includegraphics{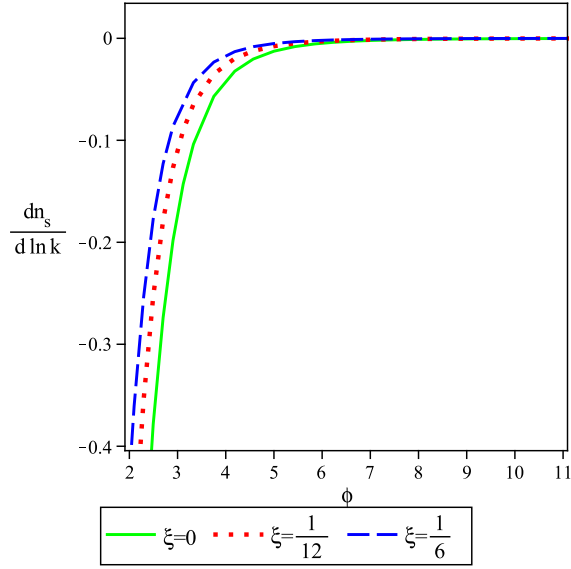}
\end{center}
\vspace{7.5 cm}
 \caption{\small { Running of the spectral index versus $\varphi$ in high-dissipation limit
 for potential  $V=V_0\varphi^2$. }}
\end{figure}

\clearpage

\end{document}